\documentclass[aps,prl,showpacs,twocolumn,superscriptaddress]{revtex4-1}
\usepackage{graphicx}
\usepackage{dcolumn}
\usepackage{bm}
\usepackage{color}

\usepackage[alsoload=hep]{siunitx}

\newunit{\atomicunit}{\textmd{a{.}u{.}}}			

\begin{document}
\title{Using circular dichroism to control energy transfer in multi-photon ionization}

\author{A.H.N.C.\ De Silva}
\affiliation{Physics Department and LAMOR, Missouri University of Science \& Technology, Rolla, MO 65409, USA}

\author{D.\ Atri-Schuller}
\affiliation{Department of Physics  and Astronomy, Drake University, Des Moines, Iowa 50311, USA}

\author{S.\ Dubey}
\affiliation{Physics Department and LAMOR, Missouri University of Science \& Technology, Rolla, MO 65409, USA}

\author{B.P.\ Acharya}
\affiliation{Physics Department and LAMOR, Missouri University of Science \& Technology, Rolla, MO 65409, USA}

\author{K.L.\ Romans}
\affiliation{Physics Department and LAMOR, Missouri University of Science \& Technology, Rolla, MO 65409, USA}

\author{ K.\ Foster}
\affiliation{Physics Department and LAMOR, Missouri University of Science \& Technology, Rolla, MO 65409, USA}

\author{O.\ Russ}
\affiliation{Physics Department and LAMOR, Missouri University of Science \& Technology, Rolla, MO 65409, USA}

\author{K.\ Compton}
\affiliation{Physics Department and LAMOR, Missouri University of Science \& Technology, Rolla, MO 65409, USA}

\author{C.\ Rischbieter}
\affiliation{Physics Department and LAMOR, Missouri University of Science \& Technology, Rolla, MO 65409, USA}

\author{N.\ Douguet}
\affiliation{Department of Physics, Kennesaw State University, Kennesaw, Georgia 30144, USA}

\author{K.\ Bartschat}
\affiliation{Department of Physics  and Astronomy, Drake University, Des Moines, Iowa 50311, USA}

\author{D.\ Fischer}
\affiliation{Physics Department and LAMOR, Missouri University of Science \& Technology, Rolla, MO 65409, USA}

\date{\today}

\begin{abstract}
Chirality causes symmetry breaks in a large variety of natural phenomena ranging from particle physics 
to biochemistry. We investigate one of the simplest conceivable chiral systems,  
a laser-excited, oriented, effective one-electron Li target. Prepared in a polarized $p$ state with $\left|m\right|=1$ in an optical trap,
the atoms are exposed to co- and counter-rotating circularly polarized femtosecond laser pulses.
For a field frequency near the excitation energy of the oriented initial state, a strong circular dichroism is observed and 
the photo\-electron energies are significantly affected by the helicity-dependent Autler-Townes splitting.  
Besides its fundamental relevance, this system is suited to create spin-polarized electron pulses with a 
reversible switch on a femtosecond timescale at an energy resolution of a few~meV.
\end{abstract}

\maketitle

Circularly polarized light exhibits handedness, a feature that gives rise to symmetry breaks in its inter\-action with matter. 
This intriguing phenomenon is well known as circular dichroism (CD), 
and it unfolds, e.g.,~as the difference in photo\-electron angular distributions (PADs) for opposite photon helicities 
in single-photon ionization of oriented diatomic molecules \cite{Jahnke2002} and even of ground-state atomic 
targets \cite{Berakdar1992,Mergel1998,Feagin2002,Hofbrucker2018}. For these non\-chiral targets, however, the 
systems of opposite photon handedness are merely mirror images of one another (\hbox{neglecting} parity-violating 
effects \cite{Loving1975,Bucksbaum1981}). This mirror symmetry is only lifted, therefore, if the target also possesses a handedness. 
Chiral molecules, i.e.,\ molecules that are not superimposable with their mirror images, are a prominent example 
of such handed targets. Their ionization by single- \cite{Boewering2001,Garcia2013,Nahon2015} or 
multi-photon \cite{Goetz2019,Lux2012,Lux2015} absorption as well as strong optical 
fields \cite{Beaulieu2018} can reveal significant dichroic asymmetries even for randomly oriented molecules. 
Such asymmetric photoreactions have far-reaching implications that could contribute to the solution of 
the long-standing puzzle of the homochirality of amino acids and sugar molecules, which are relevant 
for terrestrial life \cite{Jorissen2002, Nahon2007}.

Single atoms can also feature chirality if their orbital angular momentum is polarized along the 
projectile beam direction with a mean magnetic quantum number \hbox{$\left< m\right> \neq 0$~\cite{Mazza2014}}. In contrast to chiral molecules, 
these  systems are still superimposable with their mirror images, i.e., they do not have an intrinsic chirality. However, the atoms' 
helicity combined with an external anisotropy, e.g.\ given by the direction of an incoming photon's spin, results in a handedness, 
which sometimes is referred to as ``external chirality'' \cite{Mazza2014}.
Due to their comparably simple structure, polarized atomic targets represent benchmark systems 
for our understanding of asymmetries in the interaction of chiral light with chiral matter. 

Recent studies of atomic dichroic effects focussed, amongst other things, on fundamental aspects of magneto-optics \cite{Choi2007} 
or on the details of tunneling dynamics \cite{Herath2012,Eckart2018,Sainadh2019} and resonance-enhanced 
multi-photon ionization (REMPI) \cite{Mazza2014,Ilchen2017,GrumGrzhimailo2019}.
Circular dichroism is typically stronger in polarized atoms than in chiral molecules, because it occurs already in the 
electric dipole approximation while molecular targets require generally magnetic (or, for oriented molecules, higher-order electric) 
contributions to expose asymmetries (e.g.\ \cite{Berova2014}). For electric dipole transitions, 
the magnetic quantum number~$m$  changes by $+1$ or $-1$ for each absorbed photon of right- or left-handed circular polarization, respectively. 
Consequently, the helicity dependence in ionization of polarized atoms can be explained in terms of different 
partial waves contributing and interfering in the final state \cite{Bethe77,Thini2020}, resulting in 
dichroic asymmetries in total ionization yields and PADs.
Dichroic shifts in the photo\-electron energies, in contrast, are either completely 
absent \cite{Thini2020} or relatively small (as compared to peak widths and positions), but they can 
give insights into the structure of the dressed target atoms \cite{Ilchen2017,GrumGrzhimailo2019,Kazansky2011} 
or reveal fingerprints of atomic ring currents \cite{Eckart2018, Barth2011}.

In this Letter, we demonstrate an atomic multi-photon ionization scheme, in which circular dichroism manifests 
itself in strong and controllable shifts in the photo\-electron energy spectrum. Alkali atoms are optically 
pumped to a polarized $p$-state and subsequently ionized  by the absorption of two photons in the circularly 
polarized field of an intense femto\-second laser. A change of the relative helicity of atoms and field results in shifts 
of the photo\-electron energies by up to 40\,\% (about 100\,meV). This observation is qualitatively understood by the 
polarization-selective Autler-Townes splitting of the initial state due to its coupling to the ground state 
in the intense light field. The counter-intuitive energy dependence on the photons' polarization adds a new dimension to dichroic phenomena in photon-atom interactions and provides 
an additional dial for the quantum control of the emission of polarized electrons \cite{Hartung2016,Liu2018}. 
It can also be used to enhance the chiral response in the analysis of handed targets.

The present study contributes to the interesting and much-debated question whether photoionization 
proceeds more efficiently for the electron current density of the initial state being co-rotating or 
counter-rotating with the ionizing field. For low-intensity single-photon absorption, it is well-established 
that ionization is strongly favored in the co-rotating case \cite{Bethe77}, but the trend was found to be reversed 
in the non\-adiabatic tunnel ionization regime \cite{Barth2011,Herath2012}. For multi-photon ionization, in contrast, 
this question was not answered unambiguously.  It was found that the favored geometry swaps with increasing field 
intensity \cite{Ilchen2017,Bauer2014}. In these studies, intermediate excited states play an important role and the observed intensity dependence can be very strong \cite{GrumGrzhimailo2019} due to transient (``Freeman") resonances \cite{Freeman1987}, where 
quasi-energies of dressed intermediate states are moved in resonance, 
thereby enhancing specific REMPI channels. For the present system, in contrast, 
intermediate excited states and resonance enhancement do not substantially affect the ionization process.
Consequently, our setup represents a particularly clean manifestation of dichroic asymmetries, and it bridges the gap between the single-photon and the strong field regimes, where the details of the target structure sway the helicity dependence of the ionization yield only marginally.

In our experiment, an atomic target gas cloud was prepared in a near-resonant all-optical laser 
atom trap (AOT) \cite{Sharma2018}, where lithium atoms are cooled to temperatures of about 1\,mK. 
The wavelength of the AOT laser field is tuned near the $2s-2p$ resonance at 671\,nm ($\Delta\nu_\mathrm{AOT} < 15$\,MHz), 
resulting in a steady-state atomic excitation fraction of about 25\,\%. As shown 
earlier \cite{Sharma2018}, optical pumping in the AOT results in a high degree of atomic polarization, 
with 93\,\%  of the excited state atoms populating a single magnetic sub-level with $\left|m\right|=1$. The femto\-second 
light source is a commercially available few-cycle optical parametric chirped-pulse amplifier (OPCPA) 
system similar to the setup described in \cite{Harth2017}. It is based on a Ti:Sa oscillator providing 
the seed for two non-collinear optical parametric amplifier (NOPA) stages. For the present experiment, 
the system was configured to emit pulses with a wavelength, duration, and repetition rate 
of $665\pm 5$\,nm, 65\,fs, and 200\,kHz, respectively, and a peak power of up to $10^{12}$\,W/cm$^2$. 
The femto\-second laser beam is focused and guided through the vacuum chamber with a waist of 50\,$\mu$m 
at the target position and an angle of 10$^\circ$ with respect to the polarization direction of the 
atoms (i.e.\ the $z$-axis). Electron and recoil ion momenta are measured in coincidence in a cold-target 
recoil-ion momentum spectrometer (COLTRIMS)~\cite{Hubele2015}.

\begin{figure}
\centering
\includegraphics[width=1\linewidth]{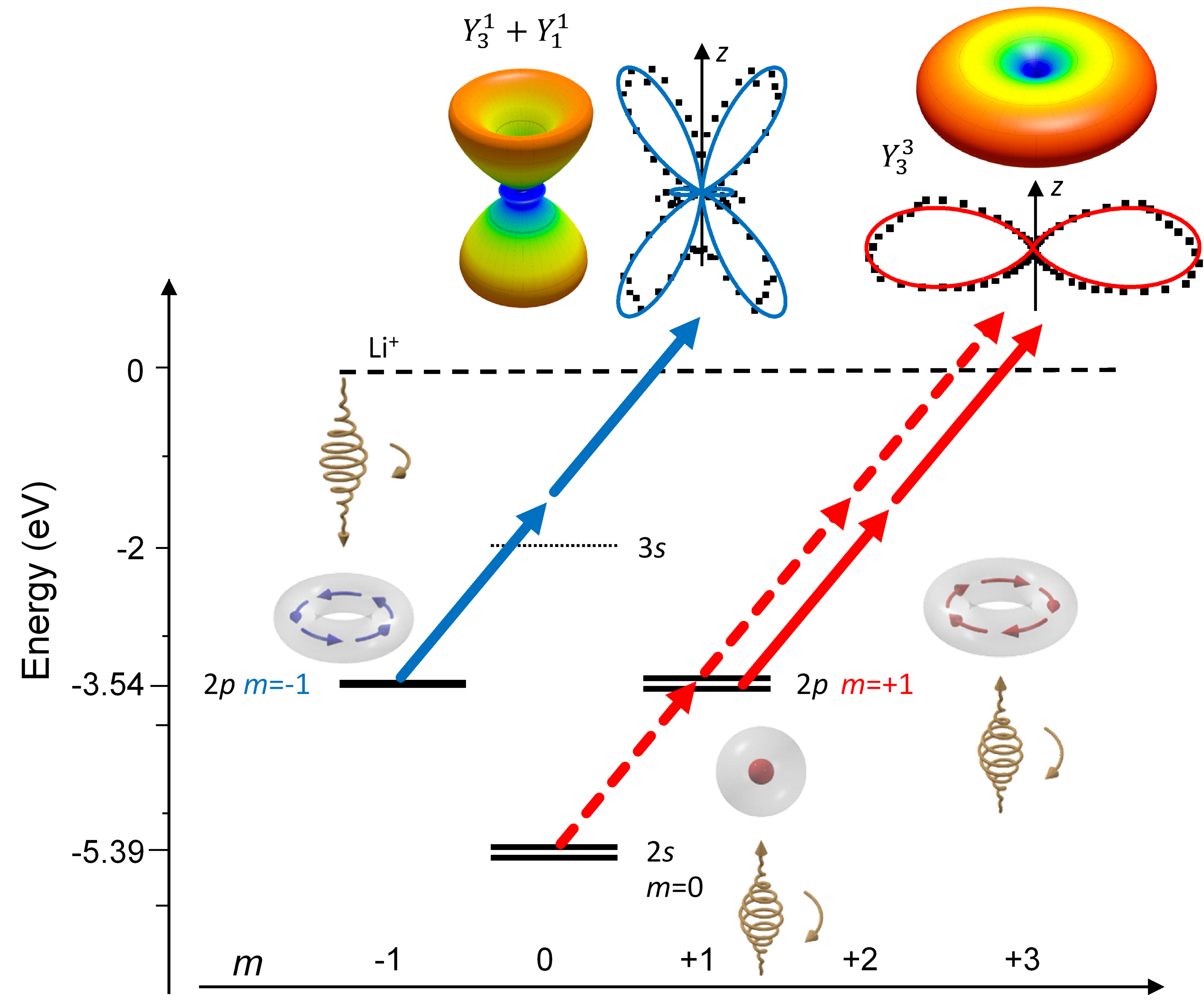}
\caption{Scheme of few-photon ionization of Li$(2p)$ and Li$(2s)$ in lowest-order perturbation theory. 
The magnetic quantum number $m$ is denoted with respect to the direction of the photon spin. 
Atomic levels undergoing Autler-Townes splitting are shown as double lines in the graph (see text). 
Measured PADs and fitted spherical harmonics are shown on the top for the counter- (left) 
and co-rotating case (right). \label{fig:IonSch}}
\end{figure}

The experimental data shown below are compared with predictions from an {\it ab initio} calculation
based on the solution of the time-dependent Schr\"odinger equation (TDSE).  For the setup considered
here, lithium can be well described as an active $(n\ell)$ valence electron above an inert He-like ($1s^2$) core.
The latter was simulated by the static Hartree potential supplemented by phenomenological
terms to simulate the core polarizability as well as exchange between the valence electron and the core.
The ideas of the method were described in~\cite{Albright_1993,CAP-book} and successfully used by~\citet{Schuricke2011}.
With a few further improvements, we obtained the ionization potentials of the $2s$ and $2p$ orbitals, as well as those of the  $n=3$ orbitals, 
to better than 1~meV of the recommended data~\cite{NIST-ASD}.  The initial state was then propagated in time
by solving the TDSE numerically~\cite{Grum_Grzhimailo_2006,PhysRevA.81.043408}.
We used an updated version of the code with the necessary modifications introduced for 
circularly polarized light described by \citet{PhysRevA.93.033402}.  The TDSE was solved in the velocity gauge and we employed a non-uniform radial grid, 
decreasing in the density of points from the region close to the nucleus (mesh $\Delta r = 0.01$\,a.u.) to the far region 
where we only need to describe slow electrons in this experiment (maximum mesh $\Delta r = 0.5$\,a.u.). 
The time step and maximum angular momentum used for the highest laser intensity were equal to $\Delta t = 0.02$\,a.u. and $\ell_{\rm max}=12$, respectively. 
Finally, we have checked that our results are converged to high accuracy.

For the conditions described above, the corresponding Keldysh paramaters are always larger than about 6.5 and the ionization process is well described in a 
multi-photon picture. The ionization pathways are depicted in Fig.~\ref{fig:IonSch} 
according to lowest-order perturbation theory. Lithium atoms in the $2s$ ground state 
are ionized by the absorption of three photons, resulting in a final-state (orbital) angular momentum 
of $(\ell,m)=(3,3)$. Note that the center frequency of the laser pulses is near the $2s-2p$ 
zero-field resonance energy with a blue-shift of $\Delta\nu_\mathrm{fs} \approx 4$\,THz (16\,meV). 
For target atoms in the excited $2p$ state, two photons suffice to promote the valence electron to 
the continuum. For the co-rotating case, the final angular momentum is identical to 
$2s$ ionization and hence given by $(3,3)$. For the counter-rotating case, on the other hand, the final magnetic 
quantum number is $m=1$, with the total angular momentum in a superposition of 
$\ell=1$ and $\ell = 3$. The different angular momenta result in  vastly different 
PADs, as illustrated in Fig.~\ref{fig:IonSch}.

\begin{figure}
\centering
\includegraphics[width=1\linewidth]{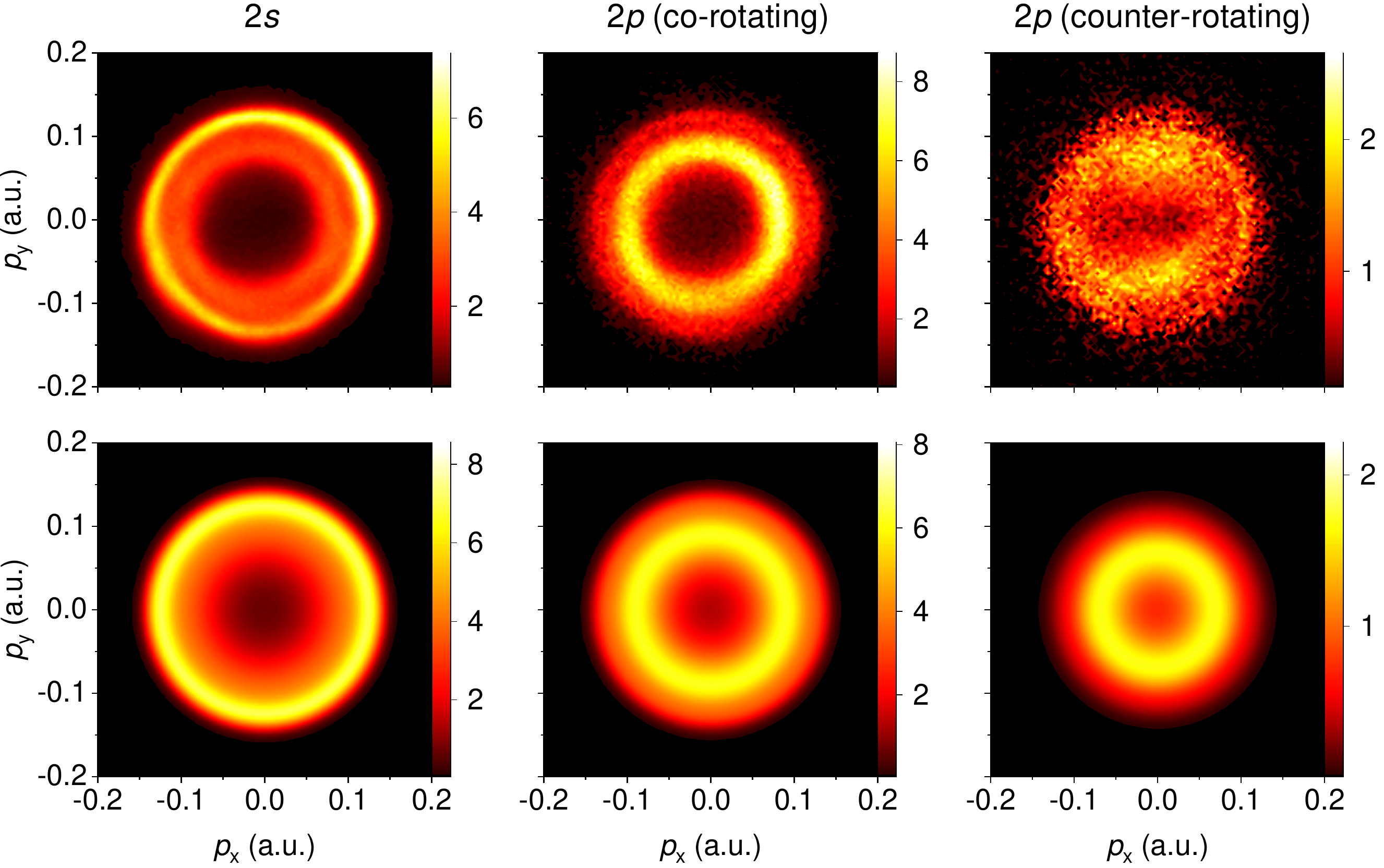}
\caption{Measured (top row) and calculated (bottom row) electron momentum distributions in the $xy$-plane 
integrated over the $z$-component for a laser field peak intensity of $0.68\times 10^{12}$\,W/cm$^2$. The initial target states are 
$2s$ (left column) and $2p$ for co-rotating (center) 
and counter-rotating (right) circular field polarizations. 
\label{fig:momXY}}
\end{figure}

Figure~\ref{fig:momXY} exhibits momentum distributions of low-energy electrons in 
the plane perpendicular to the laser beam propagation direction (the $xy$-plane) for $2s$ ionization 
as well as for co- and counter-rotating $2p$ ionization. All spectra feature ring structures 
due to the cylindrical symmetry of the systems. The diameters of the rings reflect different electron continuum energies. 
In order to account for the spatial laser field intensity distribution in the reaction volume around the focal point, 
the theoretical spectra are not calculated for a single intensity, but are weighted averages \cite{Schuricke2011} 
covering an intensity range of more than one order of magnitude with a maximum peak intensity of $0.68\times 10^{12}$\,W/cm$^2$. 
Additionally, the theoretical spectra were convolved over the experimental energy resolution of 30\,meV. 
This procedure yields excellent agreement with the experimental spectra. 
The statistical quality of the experimental data for the counter-rotating case is 
worse than for the other two situations, essentially because a relatively small 
amount of data for the $2p$ ionization is obtained by subtracting a 
large ``background" from the $2s$ initial state, which constitutes about 75\,\% of the total target density.

\begin{figure}

\centering
\includegraphics[width=1\linewidth]{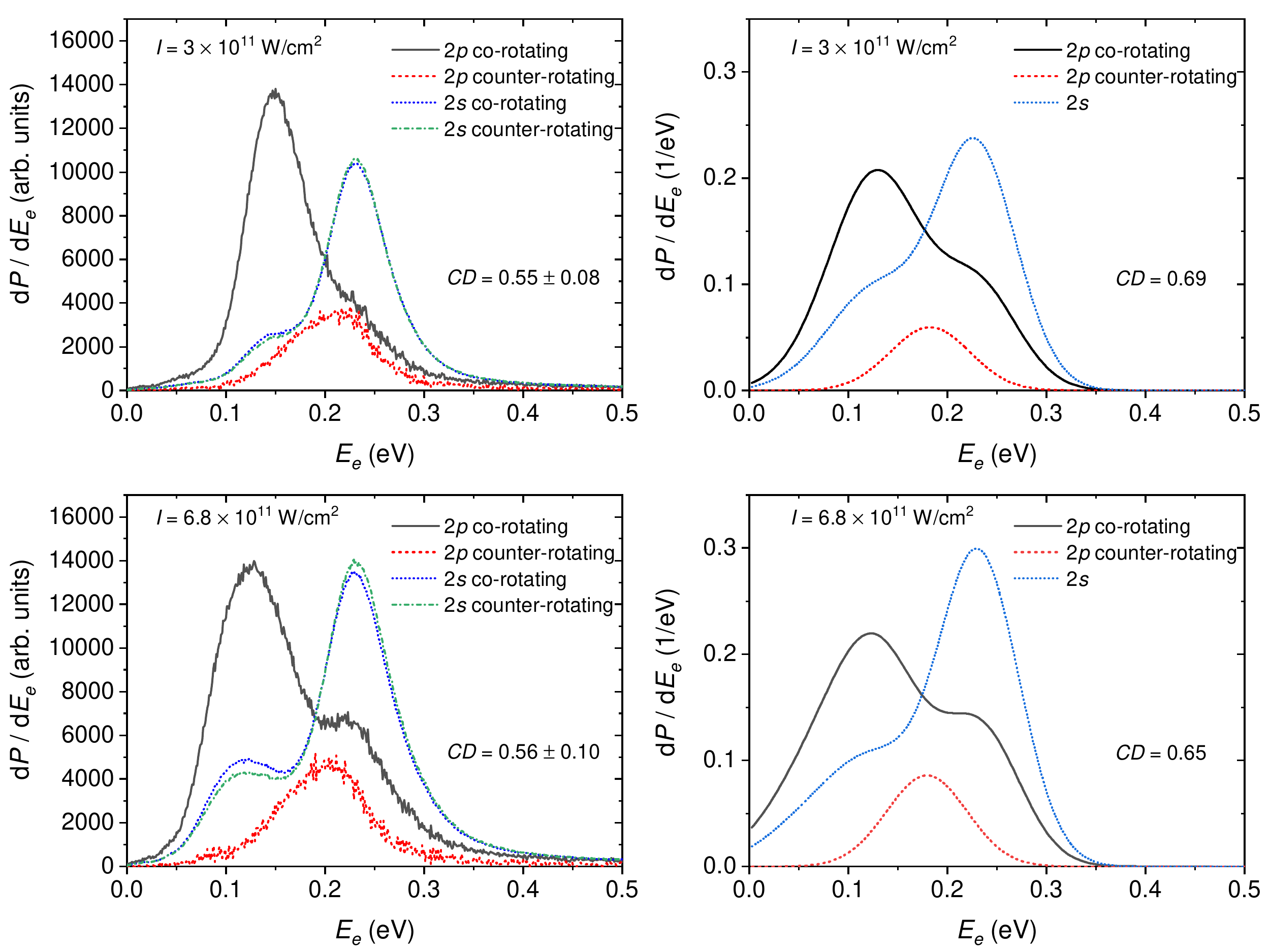}
\caption{Measured (left) and calculated (right) electron energy spectra for $2s$ and $2p$ ionization for both 
relative helicities of the circularly polarized laser field. Peak intensities $I$ and $CD$ values are given in the graphs. 
\label{fig:Ee}}
\end{figure}

The differential ionization probabilities as a function of the photo\-electron energy are plotted in Fig.~\ref{fig:Ee}. 
The energy range shown in the figure contains more than 98\,\% of the theoretical cross sections, i.e.\
contributions of higher electron energies due to above-threshold ionization (ATI) 
are small for the investigated field intensities. For ionization of the $2s$ ground state, 
the distributions for co- and counter-rotating polarization are expected to be identical 
(neglecting the spin polarization of the target atoms), which is consistent with the experimental observations. 
Remaining deviations for the two laser polarizations are attributed to systematic uncertainties, 
such as small drifts in the laser spectrum or slightly different residual ellipticities of 
the laser polarization in the two measurements.  At the lower laser intensity (top row in Fig.~3) and for 
ground-state ionization, the photo\-electron energy peak features a shoulder towards 
lower electron energies, which develops into a separate maximum with increasing laser intensity. 
A similar behavior is observed for the $2p$-state ionization for the co-rotating 
laser polarization, with the shoulder being towards the high-energy side of the main energy peak. 
In the counter-rotating situation, however, there is a single peak, whose position and general 
shape do not significantly vary with the laser intensity.

The general features of the experimental spectra are very well reproduced by our calculation,
which again has been averaged and convolved according to the method discussed above. 
However, some differences between experiment and theory persist, in particular with respect to the relative intensities of the lines.
On the theoretical side, deviations caused by model-related approximations as well as 
numerical uncertainties of the calculation are expected to be very small. 
Therefore, remaining discrepancies are more likely due to the many uncertainties 
in the experimental parameters (e.g. spectrum, spatial as well as temporal intensity distribution, 
and polarization of the laser pulses, or the polarization and population ratios of the target gas cloud), 
which are challenging to characterize precisely. 

The structure and shifts observed in the energy spectra for $2s$ ionization and for $2p$ 
ionization with co-rotating polarization can be qualitatively interpreted by the ``dressing" 
of the initial states in the photon field. The dressed-state approach is widely used to explain 
structures in photo\-electron energy spectra for multi-photon ionization of atoms and molecules
 (e.g.\  \cite{GrumGrzhimailo2019, Duncan1995}) and provides intuitive insights into the 
 physical mechanisms at play. Particularly interesting is the situation where the photon 
 field is at resonance and couples two atomic levels. The coupling splits the levels into 
 Autler-Townes doublets that are well known in atomic spectroscopy \cite{Autler1955} 
 and multi-photon ionization \cite{Sun2003,Wollenhaupt2005}. In the dressed-atom description, 
 these doublets stem from avoided crossings of the combined ``atom + photons'' states (or Floquet states) 
 at the resonant field frequency (e.g.\ \cite{Cohen-Tannoudji1996}).  Their separation 
 depends on the strength of the coupling, i.e., on the intensity of the coupling field as 
 well as the dipole moment of the atomic transition. 

In the present system, the field frequency is close to the $2s-2p$ resonance with a slight 
blue shift, splitting the two levels into doublets, which materialize as two lines (or one line with a shoulder) 
in the photo\-electron energy distributions. For the counter-rotating field, the $2p$ state does not exhibit 
the Autler-Townes splitting, because the excited initial state is not coupled to the ground state by 
radiation of opposite helicity  (see Fig.~\ref{fig:IonSch}). Generally, the evolution of dressed states 
in a femto\-second laser pulse is a time-dependent problem. Assuming the rather unrealistic case of a fully 
adiabatic evolution of the dressing, the blue-shifted radiation would cause an up-shift (down-shift) 
in energy of the initially undressed $2s$ ($2p$) state, i.e., only one level out of the respective 
doublet would be populated. Therefore, the observation of both lines of the doublets might 
generally indicate the non\-adiabaticity of the process with their relative intensities even 
providing a quantitative measure. Additionally, the electron energies are subject to ponderomotive 
shifts.  These are, however, rather small ($\leq 21$\,meV) for the present field intensities.

Quantitatively, circular dichroism is given by the difference of the relative ionization 
yields for the two photon spins.  It is defined 
as $CD=(P_+-P_-)/(P_++P_-)$, where $P_+$ and $P_-$ are the ionization probabilities 
for co- ($+$) or counter-rotating ($-$) helicities, respectively. For the two peak 
intensities shown in Fig.~\ref{fig:Ee}, the measured angle- and energy-integrated $CD$ values 
are $0.55\pm 0.08$ and $0.56\pm 0.10$, respectively. Both values are in good agreement with the theoretical predictions of~0.69 and~0.65, respectively. 
However, the measured values have significant uncertainties, which are dominated by the imprecise knowledge of the excited-state 
fraction of the atomic gas cloud ($25 \pm 3\,$\%). Contributions due to statistical errors and the cross-normalization 
procedure of the spectra for co- and counter-rotating radiation are relatively small. The given 
errors do not account for systematic effects due to the imperfect polarization of target and 
laser field, which generally are expected to shift the absolute $CD$ value slightly down. 

It is interesting to compare the present results to other recent 
studies of circular dichroism in multi-photon ionization of other atomic systems. 
Specifically, \citet{Ilchen2017} and \citet{GrumGrzhimailo2019} investigated circular 
dichroism in the double ionization of helium in an XUV-IR two-frequency field. 
In this system, the absorption of two XUV photons results in the sequential and 
resonant ionization-excitation of the target to the polarized He$^+(3p)$ state, 
which is subsequently ionized by the absorption of four or more IR photons. 
Here, the integrated $CD$ value is close to +100\% at low intensities before it 
drops and even changes sign for higher intensities. Interestingly, 
this change occurs over a very narrow intensity range. Doubling the 
intensity already suffices to bring the $CD$ value down to nearly zero.

For the present system, the overall intensity dependence of the dichroism is much weaker and appears to be more 
consistent with an earlier theoretical study considering state-prepared 
atomic hydrogen \cite{Bauer2014}. This different behavior can be understood by 
the vastly unalike preparation methods of the polarized $p$ states: 
In \cite{Ilchen2017}, the target excitation and the multi-photon ionization 
processes occur both on the same time scale in a two-frequency femto\-second radiation pulse. 
The steep drop of the dichroism is explained by the polarization-selective 
dynamic Stark shift of the He$^+(3p)$ state in the intense IR field shifting the 
XUV field and the excited target state out of resonance~\cite{GrumGrzhimailo2019}. 
In the present study, in contrast, 
the target excitation and the multi\-photon ionization processes are largely disentangled, 
as the lithium atoms are excited in the quasi-continuous, low-intensity 
(in the order of 10$^{-2}$\,W/cm$^2$) resonant field of the AOT cooling lasers 
on a much longer time scale given by the lifetime of the excited state (about 27\,ns). 
Therefore, dynamic Stark shifts of the excited $2p$ state in the femto\-second laser 
pulse do not significantly hamper the efficiency of the state preparation.

In conclusion, we calculated and demonstrated experimentally a multi-photon ionization 
scheme where strong circular dichroism occurs in the photoelectron energy distribution. 
Specifically, polarized atomic lithium in the excited $2p$ state is ionized 
by intense circularly polarized radiation of both relative helicities with a frequency near the 
$2p$ excitation energy.  If the laser electric field and the target electron current density 
counter-rotate in the same plane, the photoelectron energy spectrum exhibits a single peak at 
about $2\hbar\omega - I_P - U_P$ following simple energy conservation, with $\omega$, $I_P$, 
and $U_P$ being the field frequency, the ionization potential of the excited initial state,
and the ponderomotive energy shift, respectively. For the co-rotating case, in contrast, this energy 
relation is violated, because the $2p$ initial state and the $2s$ ground state are coupled by 
the laser field, thus resulting in the Autler-Townes splitting of both states. This effect enables to 
control photoelectron energies by the field's intensity and polarization and induce shifts that, 
in the current experiment, amount up to 40\,\% of the average continuum energy.

The multi-photon ionization scheme discussed in this Letter is ideally suited and directly 
applicable to create spin-polarized electron beams. It has been shown earlier that photoelectrons 
have a non\-vanishing spin polarization depending on their continuum energy in multi\-photon \cite{Liu2018} 
or strong-field \cite{Hartung2016} ionization of noble-gas atoms by circularly polarized light. 
Due to the state preparation of the target by optical pumping in the present scheme, not only the 
orbital angular momentum but also the spin of the single valence electron 
are aligned in the initial state \cite{Sharma2018}. Therefore, a nearly complete spin polarization of the 
photoelectrons can be expected, irrespective of their final energy. The polarization-dependent energy shift 
discussed above provides an extremely fast, femto\-second switchable dial to control the electron 
energy on a level of a few meV. This way, femtosecond spin-polarized electron pulses can be created 
with applications in electron diffraction experiments probing, e.g., ultrafast spin dynamics of magnetic domains.

\medskip
The experimental material presented here is based upon work supported by the National Science 
Foundation under Grant No.~PHY-1554776. 
We thank Thomas \hbox{Binhammer} for very helpful advice about the tunability of the laser spectrum. 
The theoretical part of this work was funded by the NSF under Grants No. PHY-1803844 (D.A.S., K.B.) and No. PHY-2012078 (N.D.), and  the XSEDE supercomputer allocation No. PHY-090031.
The calculations were carried out on Comet at the San Diego Supercomputer Center.

\end{document}